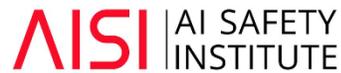

# Emerging Practices in Frontier AI Safety Frameworks

Marie Davidsen Buhl, Ben Bucknall, Tammy Masterson

*As part of the Frontier AI Safety Commitments agreed to at the 2024 AI Seoul Summit, many AI developers agreed to publish a safety framework outlining how they will manage potential severe risks associated with their systems. This paper summarises current thinking from companies, governments, and researchers on how to write an effective safety framework. We outline three core areas of a safety framework - risk identification and assessment, risk mitigation, and governance - and identify emerging practices within each area. As safety frameworks are novel and rapidly developing, we hope that this paper can serve both as an overview of work to date and as a starting point for further discussion and innovation.*

## Introduction

At the AI Seoul Summit in 2024, a number of AI developers signed on to the Frontier AI Safety Commitments, agreeing to develop a safety framework outlining how they will manage severe risks that their frontier AI systems may pose ([DSIT, 2024](#)). Since then, a research field has begun to emerge, with a diverse array of researchers from companies, governments, academia and other third-party research organisations publishing work on how to write and implement an effective safety framework. Signatories to the commitments are due to publish safety frameworks shortly, in time for the Paris AI Action Summit.

This paper summarises emerging practices – practices that appear promising and are gaining expert recognition - for safety frameworks as identified by this new research field. We draw on both the safety frameworks published so far, literature and standards on frontier AI risk management (as well as risk management more broadly), internal research by the UK AI Safety Institute, and the Frontier AI Safety Commitments themselves. The practices thus go beyond what is included in published safety frameworks and include suggestions by researchers for novel practices developers could adopt. As such, the paper provides aspirational examples for where safety frameworks might go next. At the same time, the field is still nascent and the paper is intended as a snapshot in time rather than a final and exhaustive list of practices. We hope the paper will be a useful resource for developers at all stages of development of safety frameworks, and a starting point for further research and innovation.

We organise the paper according to three core areas covered by safety frameworks: risk identification and assessment (Section 1); risk mitigation (Section 2); and governance (Section 3). Within each area, we identify components of a safety framework, and within each component, we identify emerging practices.

## Overview of emerging practices

A safety framework is a developer's plan for anticipating and managing severe risks that may arise from their frontier AI systems[1] ([DSIT, 2024](#)). It covers the three core areas:

1. **Risk identification and assessment:** Which risks are addressed by the framework and how will the developer measure if they have materialised?

2. **Risk mitigation:** Which mitigations will be implemented to reduce or respond to risks?

3. **Governance:** How will the developer make decisions based on the framework in an accountable and transparent way?

Together, these three areas form a set of "if-then" commitments (i.e. commitments about what the developer will do *if* they see certain evidence of risk), as well as commitments to support effective implementation.

We identify thirteen components of a safety framework across these three areas, based on the Frontier AI Safety Commitments (FAISCs) ([DSIT, 2024](#)). Within these components, we identify a total of 56 emerging practices, based on existing safety frameworks, best practices from other industries, and novel research. Table 1 summarises the components and emerging practices. The remainder of the paper describes them in more detail.

---

[1] As per the Frontier AI Safety Commitments, we define 'frontier AI' as highly capable general-purpose AI models or systems that can perform a wide variety of tasks and match or exceed the capabilities present in the most advanced models ([DSIT, 2024](#)).

| Area | Component | Description in the Frontier AI Safety Commitments (FAISCs) | Emerging practices |
|---|---|---|---|
| 1.Risk identification and assessment | 1.1 Risk domain identification | "Assess the risks posed by [...] frontier AI models or systems." (I) | 1. Conducting broad horizon scanning<br>2. Prioritising risk domains |
| | 1.2 Risk modelling | "[Thresholds for intolerable risk] should [...] be accompanied by an explanation of how thresholds were decided upon, and by specific examples of situations where the models or systems would pose intolerable risk." (II) | 1. Consulting with domain experts<br>2. Gathering empirical data where possible<br>3. Conducting forecasting exercises |
| | 1.3 Thresholds | "Set out thresholds at which severe risks posed by a model or system, unless adequately mitigated, would be deemed intolerable." (II) | Risk thresholds:<br>1. Pre-committing to thresholds<br>2. Collaborating with external stakeholders<br><br>Capability thresholds:<br>1. Using several tiers<br>2. Setting precise thresholds<br>3. Setting grounded thresholds<br>4. Using expert input |
| | 1.4 Model evaluation | "Assess whether [...] thresholds have been breached, including monitoring how close a model or system is to such a breach." (II) | 1. Evaluating throughout the model lifecycle<br>2. Evaluating frequently<br>3. Evaluating iteratively<br>4. Specifying the elicitation target<br>5. Using a variety of tasks<br>6. Assessing external validity<br>7. Using safety margins<br>8. Pairing targeted and open-ended evaluations |

| 2. Risk mitigation | 2.1 Deployment mitigations | "Articulate how risk mitigations will be identified and implemented to keep risks within defined thresholds, including safety [...] mitigations such as modifying system behaviours." (III) | 1. Applying "safety by design" principles<br>2. Implementing comprehensive mitigations<br>3. Building in redundancy<br>4. Covering both internal and external deployment |
|---|---|---|---|
| | 2.2 Security mitigations | "Articulate how risk mitigations will be identified and implemented to keep risks within defined thresholds, including [..] security-related risk mitigations such as [...] implementing robust security controls for unreleased model weights." (III) | 1. Applying cybersecurity standards<br>2. Building in redundancy |
| | 2.3 Mitigation evaluation | "Risk assessments should consider [...] the efficacy of implemented mitigations." (I) | 1. Setting specific targets<br>2. Evaluating robustly<br>3. Evaluating comprehensively<br>4. Using safety margins<br>5. Re-evaluating regularly |
| 3. Governance | 3.1 Conditions for safe development and deployment | "Set out [...] processes to further develop and deploy their systems and models only if they assess that residual risks would stay below the thresholds." (IV) | 1. Covering both internal and external deployment<br>2. Using safety cases |
| | 3.2 Emergency procedure | "Set out explicit processes they intend to follow if their model or system poses risks that meet or exceed the pre-defined thresholds." (IV) | 1. Communicating plans with key stakeholders upfront<br>2. Conducting simulation exercises<br>3. Coordinating with governments |
| | 3.3 Ongoing monitoring | "Continually invest in advancing their ability to implement [the commitments], including risk assessment and identification, | 1. Tracking key risk indicators<br>2. Investing in incident analysis<br>3. Creating bug bounty programs<br>4. Iterating on the safety framework |

| | | thresholds definition, and mitigation effectiveness." (V) | |
|---|---|---|---|
| | 3.4 Internal processes, roles, and resources | "Adhere to the commitments [...] including by developing and continuously reviewing internal accountability and governance frameworks and assigning roles, responsibilities and sufficient resources to do so." (VI) | 1. Defining roles<br>2. Assigning ownership at the leadership level<br>3. Appointing a chief risk officer (CRO)<br>4. Implementing multiple lines of defence<br>5. Allocating resources, including human capital<br>6. Promoting a risk culture<br>7. Establishing internal reporting channels<br>8. Implementing internal transparency |
| | 3.5 External scrutiny | "Explain how, if at all, external actors, such as governments, civil society, academics, and the public are involved in the process of assessing the risks of [...] AI models and systems, the adequacy of their safety framework [...], and their adherence to that framework." (VIII) | 1. Commissioning pre-deployment third party evaluations<br>2. Granting third party evaluators sufficient access<br>3. Giving third party evaluators sufficient time<br>4. Safe-harbouring good faith safety research<br>5. Commissioning compliance audits |
| | 3.6 External transparency | "Provide public transparency on the implementation of the [commitments]." (VII) | 1. Publishing the safety framework<br>2. Publishing system cards<br>3. Sharing more detailed reports with governments and trusted actors<br>4. Informing key stakeholders if risks arise |

Table 1: Summary of safety framework components and emerging practices

# 1: Risk identification and assessment

The first area covered by a safety framework is risk identification and assessment. This answers the question: Which risks are addressed by the framework and how will the developer measure if they have materialised?

Based on the Frontier AI Safety Commitments, we identify four components of a safety framework within this area: Risk domain identification (1.1), risk modelling (1.2), thresholds (1.3), and model evaluation (1.4).

## Component 1.1: Risk domain identification

### Description

Frontier AI Safety Commitment I includes a commitment to "assess the risks posed by [...] frontier AI models or systems." The first step is to identify the potential risks frontier AI systems may pose and, since safety frameworks typically only a subset of risk (e.g. severe risks), prioritise which risks to focus on. Safety frameworks typically include a list of covered risk.

Risk domain identification can be conducted iteratively with risk modelling (1.2), alternating between prioritising risk domains for further investigation and conducting increasingly detailed modelling.

### Emerging practices

1. **Conducting broad horizon scanning:** It can be useful to first produce a longlist of possible risk domains. This can help systematise risk identification and minimise the risk of missing key risk domains. Developers can identify potential risk domains by referring to existing literature (e.g. Slattery et al., 2024; Bengio et al., 2024). Developers can also use techniques for structured brainstorming, such as risk taxonomies and typologies (IEC, 2019).

2. **Prioritising risk domains:** Given the general-purpose nature of frontier AI systems, it is challenging to assess all possible use cases and risks (Anwar et al., 2024). As such, safety frameworks often focus on a shortlist of priority risk domains. Risks from the longlist generated above can be prioritized based on criteria such as scale, severity, and likelihood; the marginal contribution of the AI system; speed and controllability; the extent to which the risk can be effectively mitigated by the frontier AI developer; and urgency (if some scenarios can be expected to materialize sooner, they can serve as "early warning signs" of other scenarios). For an example of explicit risk prioritisation, see Karnofsky (2024).

### Examples from published safety frameworks

Risk domains covered by currently published safety frameworks include: CBRN (Anthropic, OpenAI, Google DeepMind, Magic, Naver), machine learning R&D (Anthropic, OpenAI, Google DeepMind, Magic), autonomy-related risks (OpenAI, Google DeepMind, Magic; Naver), cybersecurity (OpenAI, Google DeepMind, Magic), and persuasion (OpenAI).

## Component 1.2: Risk modelling

### Description

To identify risk domains (1.1) and set thresholds (1.3), risk modelling can be a particularly helpful tool. Risk modelling involves analysing the drivers and dynamics of a given risk domain, as well as assessing their likelihood and severity (IEC, 2019). Risk modelling provides a primary knowledge base which informs many components of safety frameworks, including thresholds, evaluations, and mitigations, and ensures they are grounded in the risks. The safety framework will likely not include complete risk models but may contain some outputs, such as the key risk scenarios or summaries of prioritisation criteria and rationales, as well as information about the risk modelling process.

An example risk modelling process for frontier AI could consist of the following steps:

- **Decomposing each risk domain** to generate a comprehensive longlist of potential risk scenarios. For example, for misuse risks, a decomposition could be based on different threat actors, attack types, severity, and harm vectors. This analysis can also consider relationships between risks.
- **Prioritising risk scenarios** based on criteria similar to those listed under risk domain identification (1.1).
- **Analysing priority risk scenarios**, for example by mapping out one or more causal chains or pathways to harm, analysing key bottlenecks or defences at each stage, and analysing the counterfactual contribution of AI at each stage. Such causal chains may not be purely linear, but could involve feedback loops or compounding risks. Established risk modelling techniques, such as the Fishbone method, scenario analysis, bow-tie analysis, and risk tree analysis, may be useful (IEC, 2019; Koessler et al., 2023; Campos et al., 2024).
- **Identifying key risk factors** such as model capabilities, other model attributes (e.g. robustness, steerability, interpretability, and alignment), deployment context (e.g. modalities, access levels, number and type of users), and societal defences. These risk factors can be prioritised based on criteria such as: the degree to which they are key bottlenecks; the number of scenarios in which they feature; and whether they could lead to cascading effects.

- **Iterating on the analysis** via red-teaming and expert input.

## Emerging practices

1. **Consulting with domain experts:** Domain experts, including government bodies, will often have established risk models and/or relevant private information. While they may not always be able to share all the relevant details, they can advise. It may also be useful to consult risk modelling experts.

2. **Gathering empirical data where possible:** It may be possible to gather high-quality empirical evidence to estimate the likelihood and impact of some steps in a risk model. For example, human baseline and uplift studies can provide evidence about key bottlenecks in a risk chain and the counterfactual impact of AI on those bottlenecks (UK AISI, 2024a).

3. **Conducting forecasting exercises:** Where high-quality empirical data is lagging, or as a supplement, aggregated forecasts from both domain experts and forecasting specialists ('superforecasters') can be a useful input into risk modelling. Such forecasts can estimate of the likelihood and impact of a risk domain as a whole or of key steps in a risk model. The Delphi technique is one way to conduct such exercises (IEC, 2019; Chapelle, 2018; Pritchard, 2015).

## Examples from published safety frameworks

| Safety framework | Examples of risk modelling summaries for autonomy-related thresholds in safety frameworks |
|---|---|
| OpenAI's Preparedness Framework | Critical threshold: "If the model is able to successfully replicate and survive or self-exfiltrate, controlling the model would be very difficult. Such a model might be able to also adapt to humans attempting to shut it down. Finally, such a model would likely be able to create unified, goal-directed plans across a variety of domains (e.g., from running commands on Linux to orchestrating tasks on Fiverr)." |
| Google DeepMind's Frontier Safety Framework | Critical capability level 1: "A model [able to autonomously acquire resources and using them to run and sustain additional copies of itself] could, if misused, pose difficult-to-predict and large-magnitude risks. Its adaptability would enable harmful activity via many means, and its ability to act autonomously and expand its effective capacity means its activity could be scaled significantly without being hindered by resource constraints. If misused or supported by well-equipped bad actors, such activity may be especially difficult to constrain." |

## Component 1.3: Thresholds

### Description

FAISC II includes a commitment to "set out thresholds at which severe risks posed by a model or system, unless adequately mitigated, would be deemed intolerable." Safety frameworks typically set out such thresholds and use them to define conditions for safe development and deployment (3.1).

Safety frameworks could use several types of thresholds:[2]

- **Risk thresholds** specify intolerable risk in terms of the likelihood and magnitude of harm (e.g. casualties or money lost). They can be quantitative or qualitative, and they can be specified either globally or for specific risk domains or scenarios (IEC, 2019; NIST 2023; Koessler et al., 2024; Aven 2015). The related condition for safe development and deployment could be that the developed or deployed system would not exceed the risk threshold(s). Rather than use risk thresholds to set conditions for safe development and deployment, developers could also use risk thresholds indirectly to inform capability thresholds (Koessler et al., 2024).
- **Capability thresholds** specify concrete dangerous capabilities that would be cause for concern. The related condition for safe development and deployment could be that the developed or deployed system either would not exceed capability thresholds *or* that it does exceed capability thresholds but adequate mitigations have been implemented.

Published safety frameworks to date primarily use capability thresholds to monitor risk and define conditions for safe development and deployment. Capability thresholds are useful for this purpose because they can be constructed to closely proxy many prioritized risks, are one of the most rapidly changing risk factors, and are easier for the developer to assess than many other risk factors. However, even if developers do not use explicit risk thresholds, informal consideration of risk tolerance informs other components of the framework, including capability thresholds, mitigations, and governance structures.

---

[2] Other kinds of thresholds that could, more speculatively, play a role in safety frameworks include outcome thresholds (e.g. an intermediate stage in a risk model that may lead to further harm), compute thresholds (e.g. a certain amount of training FLOP), and general capability benchmarks (e.g. a certain performance on general reasoning tasks).

## Emerging practices

While no currently published safety framework sets explicit risk thresholds, they are common in other industries (Marhavilas & Koulouriotis, 2021; Flamberg et al., 2016; CCPS, 2009). Some best practices for risk thresholds that may be applied to safety frameworks include:

1. **Pre-committing to thresholds:** Setting thresholds ahead of time can help developers set targets for risk assessments, capability evaluations, and safeguards, as well as provides assurance that the thresholds will not be retroactively fit above the risk assessment. Nonetheless, risk thresholds will likely need to be iterated and updated over time (ISO, 2018).

2. **Collaborating with external stakeholders:** Since risk thresholds are a normative question in the interest of all of society, they are typically set via a multi-stakeholder process and often by regulators (ISO, 2018).

Emerging practices for capability thresholds include:

1. **Using several tiers:** Several tiers of capability thresholds can be set, with corresponding risk mitigations (Anthropic, 2024; Google DeepMind, 2023; OpenAI, 2023). Early warning thresholds – which specify capabilities that are not yet considered dangerous but could indicate that models with dangerous capabilities may be developed soon – can provide a margin of error and give advance warning to prepare necessary mitigations.

2. **Setting precise thresholds:** Making capability thresholds more precise makes it easier and more objective to measure if the threshold has been crossed (METR, 2023). For example, capability thresholds can be defined based on the capabilities of specific threat actors, with a set amount of resources, against a set type of target. In some cases – especially for lower-level thresholds – it may be possible to tie thresholds to specific experimental results.

3. **Setting grounded thresholds:** Capability thresholds will be better proxies if they are grounded in detailed risk modelling. Having an in-depth understanding of the dynamics of risk will likely help identify the most relevant capabilities and set thresholds at an appropriate level. One systematic way to set capability thresholds is to base them on risk thresholds – i.e. capability thresholds can be set as the level of capabilities where the developer expect that the system may exceed the risk threshold(s) absent further mitigations (Koessler et al., 2024). Similarly, adequate mitigations can be defined as the mitigations that the developer expects would keep the developed or deployed system below the risk threshold. Alternatively, capability thresholds can be set based on more informal consideration of what level of risk would be intolerable absent adequate mitigation.

4. **Soliciting expert input:** Capability thresholds can also be based on expert views on how concerning various capabilities would be, for example via Delphi studies (IEC, 2019; Chapelle, 2018; Pritchard, 2015).

## Examples from published safety frameworks

| Safety framework | Examples of capability threshold for Machine Learning R&D in safety frameworks |
|---|---|
| Google DeepMind's Frontier Safety Framework | Critical capability level 1: "Could significantly accelerate AI research at a cutting-edge lab if deployed widely, e.g. improving the pace of algorithmic progress by 3X, or comparably accelerate other AI research groups." |
| OpenAI's Preparedness Framework | High threshold: "Model can execute open-ended, novel ML tasks on a production ML codebase that would constitute a significant step on the critical path to model self-improvement (e.g., an OpenAI pull request that trains a new model with highly diverse datasets to improve codebase editing performance, using internal tooling and documentation)." |

## Component 1.4: Model evaluation

### Description

FAISC II includes a commitment to "assess whether [...] thresholds have been breached, including monitoring how close a model or system is to such a breach." If using capability thresholds, dangerous capability evaluations are a common way to assess if thresholds have been breached. Safety frameworks typically explain how the developer intends to conduct such evaluations. For example, they can specify:

- When and how frequently model capabilities will be assessed.
- Principles for how capability evaluations will be conducted (e.g. using state-of-the-art (SOTA) techniques, aiming for full elicitation).
- Specific evaluations or evaluation types the developer intends to use.
- What processes will be used to validate and verify test results, including escalating to more intensive assessment if needed.

Developers could also evaluate other model features, such as robustness, steerability, and alignment (Campos et al., 2024). However, such evaluations are early-stage.

## Emerging practices:

1. **Evaluating throughout the model lifecycle:** Risks can emerge at any point, including during training (e.g. because a system may be stolen even if not deployed) and after deployment (e.g. because a system may improve via novel post-deployment scaffolding technique). During training, capabilities can be assessed at certain checkpoints. After deployment, capabilities can be assessed before model updates or new system features, at regular intervals, and in response to significant novel information (e.g. new scaffolding techniques or an incident) (METR, 2024a; METR, 2025; Stein & Dunlop, 2024).

2. **Evaluating frequently:** Evaluating frequently helps catch jumps in capabilities, which can emerge suddenly and non-linearly. In published safety frameworks, the evaluation frequency ranges from every 2x (OpenAI, 2024) to every 6x increase in effective compute (Google DeepMind, 2024), and every 3 months (Google DeepMind, 2024) to every 6 months (Anthropic, 2024). If the recent trend of increasing inference time compute continues, it might be useful to consider factoring this in as well (Snell et al., 2024; Erdil, 2024).

3. **Evaluating iteratively:** Developers can use lower-effort evaluations, such as automated benchmarking on held-out testing datasets, as a "first pass" filter to assess if further evaluations are necessary. A well-designed "first pass" evaluation sets tasks where failing (or performing worse than some pre-specified level) clearly rules out crossing capability thresholds. When "first pass" evaluations trigger, developers can then use more resource-intensive approaches, such as manual red-teaming, agent-based evaluations, supervised fine-tuning, and human uplift studies, to thoroughly assess if a model crosses capability thresholds (UK AISI, 2024b).

4. **Specifying the elicitation effort:** Model evaluation results are sensitive to the prompting techniques used and the resources invested in fine-tuning and scaffolding to further elicit model capabilities (UK AISI, 2024b; METR, 2024b). As such, it can be useful to pre-specify the level of effort the developer plans to invest. For example, safety frameworks could commit to full elicitation (METR, 2024a) or elicitation effort that is at least the level equivalent to the capabilities of relevant threat actors (UK AISI, 2024b). This may require conducting evaluations on models with access to certain tools or even on 'proto-agents' built for testing purposes (METR, 2024c).

5. **Using a variety of tasks:** A wide range of concrete capabilities within a given domain may contribute to risk (e.g. in the cyber domain, AI systems could cause risk by finding novel vulnerabilities or by increasing efficiency via automating parts of the attack chain). Performance across different tasks can vary, so it can be useful to test a comprehensive suite of relevant capabilities (UK AISI, 2024b),

6. **Assessing external validity:** In some cases, it may be feasible to conduct evaluations that closely mimic the threat model in question (e.g. human uplift evaluations). In other cases, developers will rely on evaluations that assess performance on proxy tasks. It is useful to consider potential discrepancies between the evaluation setting and real-world settings when interpreting evaluation results. Developers can also provide evidence for the relevance of proxy tasks (UK AISI, 2024b; Jones et al., 2024). One special case to consider is sandbagging: The possibility that a model is strategically underperforming on evaluations (Weij et al., 2024). While unlikely to be a serious possibility for current systems, developers may want to consider when and how their evaluations need to account for this possibility (Goemans et al., 2024).

7. **Using safety margins:** When interpreting test results, developers can include a safety margin to account for the risk of evaluations underestimating true model capabilities (METR, 2023; Alaga et al. 2024).

8. **Pairing targeted and open-ended evaluations:** Developers can complement targeted assessments on tasks "back-chained" from capability thresholds, with more open-ended and exploratory evaluations aiming to identify novel capabilities, vulnerabilities, and risks. These are often crowd-sourced to a team of external red-teamers (Shevlane et al., 2023; Ahmad et al., 2024; Anthropic, 2024).

## Examples from published safety frameworks

| Safety framework | Examples of model evaluations in safety framework |
|---|---|
| Anthropic's Responsible Scaling Policy | "Preliminary assessment: We will routinely test models to determine whether their capabilities fall sufficiently far below the Capability Thresholds such that we are confident that the ASL-2 Standard remains appropriate. [...]<br><br>Comprehensive assessment: For models requiring comprehensive testing [...]:<br>1. **Threat model mapping**: For each capability threshold, make a compelling case that we have mapped out the most likely and consequential threat models: combinations of actors (if relevant), attack pathways, model capability bottlenecks, and types of harms. [...]<br>2. **Evaluations**: Design and run empirical tests that provide strong evidence that the model does not have the requisite skills; explain why the tests yielded such results; and check at test time that the results are attributable to the model's capabilities rather than issues with the test design. [...] |
| | 3. **Elicitation**: Demonstrate that, when given enough resources to extrapolate to realistic attackers, researchers cannot elicit sufficiently useful results from the model on the relevant tasks [...] |
| | 4. **Forecasting**: Make informal forecasts about the likelihood that further training and elicitation will improve test results between the time of testing and the next expected round of comprehensive testing. |
| Google DeepMind's Frontier Safety Framework | "**Evaluating frontier models:** [...] we will define a set of evaluations called "early warning evaluations," with a specific "pass" condition that flags when a CCL may be reached before the evaluations are run again. We are aiming to evaluate our models every 6x in effective compute and for every 3 months of fine-tuning progress. [...]." |

# 2: Risk mitigation

The second area covered by a safety framework is risk mitigation. This answers the question: Which mitigations will be implemented to prevent or respond to risks materialising?

Based on the Frontier AI Safety Commitments, we identify three components of a safety framework within this area: Deployment mitigations (2.1), security mitigations (2.2), and mitigation evaluation (2.3).

## Component 2.1: Deployment mitigations

### Description

FAISC III includes a commitment to "articulate how risk mitigations will be identified and implemented to keep risks within defined thresholds, including safety and security-related risk mitigations such as modifying system behaviours and implementing robust security controls for unreleased model weights." One class of mitigations typically covered by safety frameworks is deployment mitigations, which aim to prevent (externally or internally) deployed systems from causing harmful outcomes. Such harmful outcomes can occur either due to misuse or by accident (including due to autonomous actions by the system). Safety frameworks can outline which deployment mitigations the developer expects to implement for systems at certain capability levels.

Deployment mitigations can include *system-level safeguards* that aim to prevent malicious actors from utilising dangerous capabilities (e.g. refusal training, input-output classifiers, machine unlearning, or data filtering) or prevent the model from taking harmful actions in a wide array of environments (e.g. emerging control protocols, robustness or alignment techniques). It can also include *access safeguards* that aim to prevent malicious users from accessing the system at all (e.g. usage monitoring, customer verification and vetting, or bans of suspicious accounts).

### Emerging practices

1. **Applying "safety by design" principles:** Mitigating deployment risk does not mean waiting until after training to implement measures, but includes measures built into the system design and training process, such as data filtering (CISA, 2013).
2. **Implementing comprehensive mitigations:** Specific mitigations may not be effective for all types of user interactions or deployment scenarios. For example, a system-level safeguard may not be effective if the system is accessed via third-party applications or combined with certain scaffolding, or if the user has fine-tuning access. There is value in considering a range of scenarios when planning mitigations and from taking measures to ensure that downstream developers maintain adequate mitigations.

3. **Building in redundancy:** To increase robustness, developers can aim to avoid single points of failure and instead use a "defense-in-depth" strategy, i.e. implementing multiple mitigations addressing the same threat model or failure mode (Alaga et al., 2024).

4. **Considering both internal and external deployment:** Different mitigations may be suitable for internal and external deployment due to different underlying risk models. As such, developers may want to separately specify which mitigations they plan to implement for internal use (e.g. monitoring and logging of interactions, limited affordances, staff training).

## Examples from published safety frameworks

Most published safety frameworks do not currently commit to specific deployment mitigations but rather specify criteria that the mitigations must meet. These are covered under mitigation evaluation (2.3).

## Component 2.2: Security mitigations

## Description

Another class of mitigations typically covered by safety frameworks – and referenced in FAISC III – is security mitigations, which aim to prevent the theft, leak, or self-exfiltration of models. Specifically, security mitigations aim to protect model weights and other sensitive information that may enable an adversary to build a similarly capable model. Safety frameworks can outline which security mitigations the developer expects to implement for systems at certain capability levels.

As there is a risk of model theft, leak, or self-exfiltration as soon as a model is developed, safety frameworks typically commit to putting in place the necessary security mitigations before a model is *developed*, rather than before it is deployed.

## Emerging practices

1. **Applying cybersecurity standards:** There are many existing cybersecurity standards that developers can apply to the protection of model weights, such as NCSC's guidelines on secure model development (NCSC, 2024) and DSIT's draft AI cyber security Code of Practice (DSIT, 2024). Developers can also refer to RAND's state-of-the-art overview of security measures specifically focused on model weights (Nevo et al., 2024).

2. **Building in redundancy:** As with deployment safeguards, developers can aim to implement a "defense-in-depth" strategy.

## Examples from published safety frameworks

| Safety framework | Examples of security mitigations in safety frameworks |
|---|---|
| OpenAI's Preparedness Framework | "If we reach (or are forecasted to reach) at least "high" pre-mitigation risk in any of the considered categories we will ensure that our security is hardened in a way that is designed to prevent our mitigations and controls from being circumvented via exfiltration (by the time we hit "high" pre-mitigation risk). [...]. This might require:<br>• increasing compartmentalization [...]<br>• deploying only into restricted environments [...]<br>• increasing the prioritization of information security controls." |
| Google DeepMind's Frontier Safety Framework | Level 1: "Limited access to raw representations of the most valuable models, including isolation of development models from production models. [...]"<br><br>Level 2: "Changes to ML platforms and tools to disallow unilateral access to raw model representations outside the core research team, with exceptions granted on the basis of business need."<br><br>Level 3: "Models can be accessed only through high-trust developer environments (HTDE), hardened, tamper-resistant workstations with enhanced logging."<br><br>Level 4: "Minimal trusted computing base (TCB). TPUs with confidential compute capabilities. Dedicated hardware pods for training and serving high-value models." |

## Component 2.3: Mitigation evaluation

## Description

FAISC I includes a commitment to conducting risk assessments which consider "the efficacy of implemented mitigations", and FAISC III states that mitigations should "keep risks within defined thresholds". To understand the impact of mitigations on risks, it is useful to evaluate the effectiveness of mitigations, not only in the abstract, but also specific mitigations as implemented for specific systems. This is especially the case because it can be difficult to predict what mitigations will be sufficient for a future system, given that both mitigations and our understanding of frontier AI are rapidly developing. As such, safety frameworks can commit to conducting evaluations of the mitigations implemented for specific systems. Safety frameworks can explain details such as:

- When safeguards will be evaluated, including a specification of which updates to the model or the mode of deployment will require new safeguards evaluations.
- Targets for mitigation effectiveness (e.g. a certain robustness rate, or making a certain capability inaccessible to a certain actor).
- Principles for how mitigation evaluations will be conducted (e.g. using SOTA techniques).
- Specific evaluations or evaluation types the developer intends to use.
- Processes for continuously monitoring mitigation effectiveness, including frequency of post-deployment mitigation evaluations.

## Emerging practices

1. **Setting specific targets:** Being precise about what standards mitigations aim to meet makes testing easier and reduces subjectivity. These targets can be grounded risk modelling (1.2) or based on risk thresholds (1.3). For example, a target for system-level safeguards against cyber misuse risk could be: "A technical non-expert would not be able to elicit expert-level vulnerability discovery capability within two weeks and with a budget of £1000."

2. **Evaluating robustly:** As with capability elicitation, it can be helpful to consider what level of effort is necessary to provide robust evidence of mitigation effectiveness. For example, the developer can use red-teams that are as capable and well-resourced as relevant threat actors. Developers can also aim to use multiple types of evaluations, where feasible, to increase robustness.

3. **Evaluating comprehensively:** Mitigations that are effective in one context – for example, one deployment form or one language – may not be effective in another. As such, developers can assess mitigations in all contexts in which they will be deployed (e.g. API with fine-tuning, API with internet access and tool use, etc.) and assess if mitigations remain effective across natural languages.

4. **Using safety margins:** As with capability evaluations, developers can include a safety margin to account for the risk of overestimating effectiveness ([Alaga et al. 2024](#)).

5. **Re-evaluating regularly:** Mitigation effectiveness may change during development and deployment (e.g. new jailbreaks might be discovered), or new information may appear (e.g. about safeguard vulnerabilities). To address this, developers can regularly re-assess mitigation effectiveness, at regular intervals and/or in response to pre-specified triggers such as a certain incident rate. Developers can also commit to re-assessing mitigation effectiveness when changing or expanding deployment modalities significantly.

## Examples from published safety frameworks

| Safety framework | Examples of mitigation evaluations in safety frameworks |
|---|---|
| Google DeepMind's Frontier Safety Framework | Level 1: "Application, where appropriate, of the full suite of prevailing industry safeguards targeting the specific capability, including safety fine-tuning, misuse filtering and detection, and response protocols. Periodic red-teaming to assess the adequacy of mitigations."<br><br>Level 2: "A robustness target is set based on a safety case considering factors like the critical capability and deployment context. Afterwards, similar mitigations as Level 1 are applied, but deployment takes place only after the robustness of safeguards has been demonstrated to meet the target." |
| OpenAI's Preparedness Framework | **"Evaluating post-mitigation risk:** To verify if mitigations have sufficiently and dependently reduced the resulting postmitigation risk, we will also run evaluations on models after they have safety mitigations in place, again attempting to verify and test the possible "worst known case" scenario for these systems. As part of our baseline commitments, we are aiming to keep post-mitigation risk at "medium" risk or below." |

# 3: Governance

The third area covered by a safety framework is governance. This answers the question: How will the developer make decisions based on the framework in an accountable and transparent way?

Based on the Frontier AI Safety Commitments, we identify three components of a safety framework within this area: Conditions for safe development and deployment (3.1), emergency procedure (3.2), ongoing monitoring (3.3), internal processes, roles, and resources (3.4), external scrutiny (3.5), and external transparency (3.6).

## Component 3.1: Conditions for safe development and deployment

### Description

FAISC IV includes commitments to set out "processes to further develop and deploy their systems and models only if they assess that residual risks would stay below the thresholds" and to "not [...] develop or deploy a model or system at all, if mitigations cannot be applied to keep risks below the thresholds". As such, safety frameworks typically specify conditions for safe development and deployment – with reference to the thresholds – as well as the process for determining if these conditions have been met.

On the substantive side (i.e. setting the conditions themselves), safety frameworks can include:

- An affirmation of the developer's commitment to only develop or deploy systems that do not pose risk deemed to be intolerable, unless the risk has been adequately mitigated. If using risk thresholds, this could be a commitment to only develop or deploy systems that remain below risk thresholds. If using capability thresholds, this could be a commitment to only develop and deploy systems that either do not cross capability thresholds or for which adequate mitigations have been implemented (security mitigations in the case of development; deployment mitigations in the case of deployment).
- Any conditions under which the commitment can be bypassed (e.g. if the developer believes the capability threshold in its safety framework is too low; or if a rigorous cost-benefit analysis finds that the benefits of developing or deploying the model outweighs risks – if such cost-benefit analysis has not already been used to set the thresholds).
- Additional commitments for systems that fail to meet the conditions for safe development and deployment, such as commitments to not develop other similarly capable models, reallocate staff to certain tasks, inform key stakeholders, or heavily restrict internal access to the model (METR, 2023).

On the procedural side, safety frameworks can include:

- Any assessment or documentation that must be produced to support a decision (e.g. a risk assessment, capability report or safety case).
- Any review or red-teaming process said assessment document must go through.
- Final decision-maker and owner.
- Any special requirements for deciding to bypass the commitment (e.g. board approval).

## Emerging practices

1. **Covering both internal and external deployment:** Safety frameworks typically cover both internal and external deployment. It can be useful to confirm this explicitly, especially if using different procedures to approve internal and external deployment.
2. **Making a safety case:** Safety cases – a structured argument that a system is safe in a given context – are emerging as a framework for bringing various evidence sources together in an overall risk assessment (Clymer et al., 2024; Buhl et al., 2024; Irving et al., 2024). They may be a particularly useful way to assess if the conditions for safe development and deployment have been met, as well as to present information to key decision-makers, due to their comprehensive and explicit reasoning. Ideally, a safety case would be gradually developed and updated throughout the system lifecycle and used to inform decisions such as what evaluations to run, what mitigations to implement, and how to respond to an incident.

## Examples from published safety frameworks

| Safety framework | Examples of conditions for safe development and deployment in safety frameworks |
|---|---|
| Magic's AGI Readiness Policy | "If the engineering team sees evidence that our AI systems have exceeded the current performance thresholds on the public and private benchmarks listed above, the team is responsible for making this known immediately to the leadership team and Magic's Board of Directors (BOD)." |
| Anthropic's Responsible Scaling Policy | "The process for determining whether we have met the ASL-3 Required Safeguards is as follows:<br>• First, we will compile a Safeguards Report for each Required Safeguard that documents our implementation of the measures above, makes an affirmative case for why we have satisfied them and advances recommendations on deployment decisions.<br>• The Safeguards Report(s) will be escalated to the CEO and the Responsible Scaling Officer, who will (1) make the ultimate determination as to whether we have satisfied the Required Safeguards and (2) decide any deployment-related issues.<br>• [...] For high-stakes issues [...], the CEO and RSO will likely solicit internal and external feedback on the report prior to making any decisions.<br>• If the CEO and RSO decide to proceed with deployment and training, they will share their decision–as well as the underlying Capability Report, internal feedback, and any external feedback–with the Board of Directors and the Long-Term Benefit Trust before moving forward.<br>• [Safeguards] will be revisited and re-approved at least annually [...]." |

## Component 3.2: Emergency procedure

### Description

FAISC IV includes a commitment to "set out explicit processes [the developer] intend[s] to follow if their model or system poses risks that meet or exceed the pre-defined thresholds". This is partly addressed by setting out a commitment to not develop or deploy systems that cross thresholds unless adequate mitigations are in place (3.1). However, despite their best efforts to prevent it, a developer could inadvertently develop or deploy a system without sufficient mitigations in place to prevent severe harm. example, a developer could underpredict capability improvements from scaling or post-deployment scaffolding improvements, or if mitigations could cease to be effective after deployment. To prepare for such scenarios, safety frameworks can explain how the developer will respond in a situation where the conditions for safe development and deployment have been or might have been accidentally violated. For example, the safety framework can include:

- Outline of emergency scenarios and planned emergency responses, such as rapid incident investigation, reducing access to the model or, in the most extreme cases, model shutdown.
- Commitments to alert key stakeholders such as relevant government agencies.
- Information about how the developer will test and communicate its emergency plans ahead of time.

### Emerging practices

1. **Communicating plans with key stakeholders upfront:** Developers can prepare employees and communicate plans with users and downstream developers, especially those who may use the system for critical functions. This can make it easier and less costly to implement emergency measures if and when needed (O'Brien et al., 2023).
2. **Conducting simulation exercises:** Conducting periodic simulation exercises can help stress test emergency playbooks and prepare staff (WHO, 2017).
3. **Coordinating with governments:** Governments may have key roles to play in responding to emergencies, so it may be beneficial to coordinate response plans with relevant government agencies.

### Examples from published safety frameworks

| Safety framework | Examples of emergency procedures in safety frameworks |
| --- | --- |
| Anthropic's Responsible Scaling Policy | "In any scenario where we determine that a model requires ASL-3 Required Safeguards but we are unable to implement them immediately [...]:<br><br>• **Interim measures**: The CEO and Responsible Scaling Officer may approve the use of interim measures that provide the same level of assurance as the relevant ASL-3 Standard but are faster or simpler to implement. [...]<br>• **Stronger restrictions**: In the unlikely event that we cannot implement interim measures to adequately mitigate risk, we will impose stronger restrictions. In the deployment context, we will de-deploy the model and replace it with a model that falls below the Capability Threshold. [...] In the security context, we will delete model weights. [...]." |
| OpenAI's Preparedness Framework | "**Safety drills**: [...] the SAG will call for safety drills at a recommended minimum yearly basis." |

## Component 3.3: Ongoing monitoring

### Description

FAISC V includes a commitment to "continually invest in advancing their ability to implement [the commitments], including risk assessment and identification, thresholds definition, and mitigation effectiveness. This should include processes to assess and monitor the adequacy of mitigations, and identify additional mitigations as needed to ensure risks remain below the pre-defined threshold." This commitment covers two areas: (1) Ongoing monitoring of individual frontier AI systems – including their capabilities, the effectiveness of safeguards, and their impact – and (2) Continuous updating of the safety framework itself. Safety framework can cover areas such as:

- Regular cadence or triggers for continuous model and mitigation evaluations, as covered in under model evaluation (1.4) and mitigation evaluation (2.3).
- Incident monitoring, including automated and human systems for incident detection, logging, triaging, response, documentation, and analysis.
- Tracking of novel emerging risk domains or changes to already identified risk domains (e.g., increases or decreases in societal resilience that significantly affect the magnitude of risk).
- Plans for further research and staying up-to-date with best practices in safety frameworks and capability and mitigations evaluations.
- Cadence and process for updating the safety framework.

## Emerging practices

1. **Tracking key risk indicators:** For safety-critical autonomous systems, such as autonomous vehicles, it is common to identify key risk indicators (such as incident rates and component failure rates), track these after deployment, and continuously update risk assessments accordingly (Koopman, 2022). Developers could experiment with a similar approach, using metrics such as safeguard failure rate as observed by monitoring or bug bounty fault finding rates.

2. **Investing in incident analysis:** There are often valuable lessons to be learned from the causes, frequency, and impacts of incidents. Developers can conduct incident analysis themselves, as well as support external researchers by sharing information (Stein & Dunlop, 2024). One way to support external research can be to log incidents in a shared database (McGregor, 2021).

3. **Creating bug bounty programs:** In cybersecurity, it is common to offer "bug bounties", i.e. financial prizes to individuals who identify security vulnerabilities. Similar programs could help AI developers monitor capabilities and safeguard effectiveness (Levermore, 2023; OpenAI, 2024).

4. **Iterating on the safety framework:** Safety frameworks are a novel concept, and practices are rapidly developing. As such, it can be useful to update safety frameworks regularly. If parts of the safety framework are not fully specified, the developer can provide assurance by committing to a certain timeline. For example, Anthropic commits to developing ASL-4 Capability Thresholds before deploying any model that requires ASL-3 Safeguards (Anthropic, 2024); and Magic commits to halting development of models that cross certain benchmarks until it has developed dangerous capability evaluations (Magic, 2024).

## Examples from published safety frameworks

| Safety framework | Examples of ongoing monitoring in safety frameworks |
|---|---|
| Anthropic's Responsible Scaling Policy | "**Rapid remediation:** Show that any compromises of the deployed system, such as jailbreaks or other attack pathways, will be identified and remediated promptly enough to prevent the overall system from meaningfully increasing an adversary's ability to cause catastrophic harm. Example techniques could include rapid vulnerability patching, the ability to escalate to law enforcement when appropriate, and any necessary retention of logs for these activities implementation challenges we encounter." [ASL-3 Deployment Standard] |
| OpenAI's Preparedness Framework | "We will also, in cooperation with other teams (e.g., Safety Systems), develop monitoring and investigative systems. This monitoring of real-world misuse (as well as staying abreast of |

| | relevant research developments) will help us create a better picture of deployed model characteristics, and inform updates to our evaluations as necessary." |
|---|---|

## Component 3.4: Internal processes, roles, and resources

### Description

FAISC VI includes a commitment to "developing and continuously reviewing internal accountability and governance frameworks and assigning roles, responsibilities and sufficient resources to [adhere to the commitments]." In line with this, safety frameworks can outline what processes are in place to ensure the safety framework is implemented, who is responsible for doing so, and what resources will be allocated to safety framework implementation.

### Emerging practices

In setting internal accountability structures, safety frameworks can draw on the well-established field of organizational risk management. Emerging pactices from this field that may be applied to safety frameworks include:

1. **Defining roles:** It is useful to specify a risk management structure, clearly identify responsible individuals throughout the organization, and maintain internal and external transparency about said structure (ISO 2018; NIST, 2023). Having a clear and unambiguous owner and process for final decision-making (for example, an individual final decision-maker or a clear decision rule for a group) can help prevent diffusion of responsibility. Safety frameworks can specify such a structure, or it can assign implementation responsibilities to individuals within a structure established separately.

2. **Assigning ownership at the leadership level:** Clearly assigning ownership of risk management to C-suite leadership can help ensure that it remains an organiszational priority (ISO 2018; NIST, 2023). The board of directors typically has final oversight over risk decisions; as such, it can be beneficial to establish a board risk committee (NIST, 2023; Walker, 2009). These measures could also be applied to safety framework implementation.

3. **Appointing a chief risk officer (CRO):** It can be useful to appoint an employee whose primary responsibility is to ensure that the framework is implemented, such as a chief risk officer (Robinson & Ginns, 2024). The CRO typically has an advisory role, rather than being a final decision-maker or risk owner. Developers can consider ways to empower the CRO, for example by including them in leadership teams and/or boards.

4. **Implementing multiple lines of defence:** Requiring several independent internal actors to review key documentation and decisions can help prevent errors and red team judgement calls. This is true of both day-to-day decision making, major development and deployment decisions, and procedural framework implementation. For particularly high-stakes decisions, it may be valuable to ensure review by independent actors without a commercial interest in the company. Developers could be inspired by established models such as the Three Lines Model (IIA, 2020; Schuett, 2023; Robinson & Ginns, 2024).

5. **Allocating resources, including human capital:** Safety frameworks can outline how the developer will ensure that responsible individuals have the adequate resourcing, expertise, and training to implement the safety framework effectively (ISO 2018; NIST, 2023; Robinson & Ginns, 2024; Alaga et al., 2024).

6. **Promoting a risk culture:** Safety frameworks can outline ways in which leadership intends to promote a risk culture in the organisation and ensure a sense of ownership throughout, for example via communication and staff training (Robinson & Ginns, 2024).

7. **Establishing internal reporting channels:** Establishing channels by which employees can report concerns or noncompliance with the framework, as well as protections against retaliation for employees who report concerns, can help provide assurance that the framework is correctly implemented (Robinson & Ginns, 2024).

8. **Implementing internal transparency:** Ensuring that employees are well-informed about safety framework implementation can enable employees to notice and draw attention to gaps and thereby enhance safety – so long as information sharing does not compromise security. Safety frameworks can outline what information will be shared with employees, as well as when and with whom.

## Examples from published safety frameworks

| Safety framework | Examples of internal processes, roles, and resources in safety frameworks |
| --- | --- |
| OpenAI's Preparedness Framework | "Parties in the Preparedness Framework operationalization process:<br>• **The Preparedness** team conducts research, evaluations, monitoring, forecasting, and continuous updating of the Scorecard with input from teams that have relevant domain expertise.<br>• **The Safety Advisory Group (SAG), including the SAG Chair**, provides a diversity of perspectives to evaluate the strength of evidence related to catastrophic risk and recommend appropriate actions. [...]. |
| | o The members of the SAG and the SAG Chair are appointed by the OpenAI Leadership. [...]. <br> o SAG membership will rotate yearly. [...]. <br> • **The OpenAI Leadership**, i.e., the CEO or a person designated by them, serves as the default decision-maker on all decisions. <br> • **The OpenAI Board of Directors (BoD**), as the ultimate governing body of OpenAI, will oversee OpenAI Leadership's implementation and decision-making pursuant to this Preparedness Framework. [...]." |
| Anthropic's Responsible Scaling Policy | "**Transparency**: We will share summaries of Capability Reports and Safeguards Reports with Anthropic's regular-clearance staff, redacting any highly-sensitive information. [...]." <br><br> "**Noncompliance**: We will maintain a process through which Anthropic staff may anonymously notify the Responsible Scaling Officer of any potential instances of noncompliance with this policy. We will also establish a policy governing noncompliance reporting, which will (1) protect reporters from retaliation and (2) set forth a mechanism for escalating reports to one or more members of the Board of Directors in cases where the report relates to conduct of the Responsible Scaling Officer. [...]" |

## Component 3.5: External scrutiny

## Description

FAISC VIII includes a commitment to "explain how, if at all, external actors, such as governments, civil society, academics, and the public are involved in the process of assessing the risks of [...] AI models and systems, the adequacy of their safety framework [...], and their adherence to that framework." External actors can also play a key role in providing assurance and accountability, in line with FAISC VI. Safety frameworks typically outline how external actors are involved in implementing the safety framework, including, for example, by:

- Providing feedback on the safety framework itself;
- Conducting or auditing model and mitigation evaluations across the AI lifecycle;
- Recommending mitigations based on evaluation results;
- Assessing or providing feedback on key documents (such as capability reports, safeguard reports, or safety cases) ahead of key decisions (such as deployment);
- Auditing procedural safety framework implementation and safety culture.

## Emerging practices

1. **Commissioning pre-deployment third party evaluations:** It is common for developers to grant third party evaluators access to models prior to deployment to evaluate capabilities and safeguards (see e.g. OpenAI 2024, Anthropic 2024). This can include both targeted evaluations and more exploratory red-teaming. Third party evaluations can help provide additional evidence and assurance about whether the conditions for safe development and deployment have been met. Many third party auditors, such as METR and Apollo, also specialise in particular risks and can contribute domain expertise.

2. **Granting third party evaluators sufficient access:** The ability of third party evaluators to fully elicit model capabilities and evaluate mitigation effectiveness depends on what models they have access to and how they can use them. In the current model evaluation paradigm, third party evaluators benefit from access to both pre- and post-safeguard versions of the model and fine-tuning API access. They may also benefit from support from the developer's technical staff (UK AISI, 2024b; Bucknall & Trager, 2023, Casper et al., 2024).

3. **Giving third party evaluators sufficient time:** Giving third party evaluators more time to evaluate models before deployment enables more rigorous evaluations. It may increase efficiency to conduct multiple rounds of evaluations before deployment, first sharing earlier-stage or pre-safeguards models and later sharing later-stage or post-safeguards models. If most evaluations happen on models that differ from the final deployed model, developers can commit to being transparent about this and giving the third-party evaluator access to the final model to do comparative analysis (UK AISI, 2024b).

4. **Safe-harbouring good faith safety research:** After deployment, developers can continue to learn from the insights of researchers and users formally or informally evaluating capabilities and safeguards, as well as using the model to conduct safety research. Developers can consider supporting such research providing exemptions to terms of service to vetted external researchers for the purposes of studying model safety (Longpre et al., 2024).

5. **Commissioning procedural audits:** In addition to providing model-specific evidence, third parties could also assess if the developer is following the *processes* laid out in the safety framework (e.g. commitments to produce certain documentation or establish certain internal processes) via a procedural compliance audit (Robinson & Ginns, 2024).

## Examples from published safety frameworks

| Safety framework | Examples of external scrutiny in safety frameworks |
|---|---|
| OpenAI's Preparedness Framework | "**Audits:** Scorecard evaluations (and corresponding mitigations) will be audited by qualified, independent third-parties to ensure accurate reporting of results, either by reproducing findings or by reviewing methodologies to ensure soundness. [...]."<br><br>"**External access:** We will also continue to enable external research and government access for model releases. [...]." |
| Anthropic's Responsible Scaling Policy | "**Procedural compliance review:** On approximately an annual basis, we will commission a third-party review that assesses whether we adhered to this policy's main procedural commitments (we expect to iterate on the exact list since this has not been done before for RSPs)." |

## Component 3.6: External transparency

### Description

FAISC VIII includes a commitment to "provide public transparency on the implementation of the [commitments], except insofar as doing so would increase risk or divulge sensitive commercial information to a degree disproportionate to the societal benefit and "share more detailed information which cannot be shared publicly with trusted actors". In line with this, safety frameworks can outline what information about safety framework implementation will be shared with external actors.

### Emerging practices

1. **Publishing the safety framework:** Signatories of the Frontier AI Safety Commitments have committed to publishing their safety framework. Developers can provide greater transparency by ensuring that safety frameworks describe their commitments clearly and comprehensively and by sharing the rationales behind them (Alaga et al., 2024). When updating their safety framework, developers can further transparency by maintaining a changelog as well as copies of previous versions of the framework.

2. **Publishing system cards:** When releasing a system, developers typically publish a system card summarising the system design, evaluation results, mitigations, and external scrutiny during model development – while excluding commercially sensitive information. It has also been proposed that developers produce more detailed transparency reports (NIST, 2024; Bommasani et al., 2024; Mitchell et al., 2019).

3. **Sharing more detailed reports with governments and trusted actors:** Developers can share more detailed information with relevant government agencies and trusted actors such as third-party auditors, research collaborators, or downstream developers, in line with FAISC VIII. This could include internal documentation used to inform go/no-go decisions, such as reports on model and mitigation evaluations or safety cases. It could also include periodic updates such as incident analysis (Stein & Dunlop, 2024).

4. **Informing key stakeholders if risks arise:** Governments can play an important role in conducting and coordinating risk mitigation. Other actors, such as downstream developers, users applying the system in safety-critical contexts, or operators of key national infrastructure, can also take useful actions. As such, developers can commit to inform key actors of incidents, near-misses, or key risk factors such as dangerous capabilities, even if these do not warrant an acute response (Kolt et al., 2024; O'Brien et al., 2024).

## Examples from published safety frameworks

| Safety framework | Examples of external transparency in safety frameworks |
|---|---|
| Google DeepMind's Frontier Safety Framework | "We are exploring internal policies around alerting relevant stakeholder bodies when, for example, evaluation thresholds are met, and in some cases mitigation plans as well as post-mitigation outcomes." |
| Anthropic's Responsible Scaling Policy | "**Public disclosures**: We will publicly release key information related to the evaluation and deployment of our models (not including sensitive details). These include summaries of related Capability and Safeguards reports when we deploy a model as well as plans for current and future comprehensive capability assessments and deployment and security safeguards. [...].<br><br>"**U.S. Government notice:** We will notify a relevant U.S. Government entity if a model requires stronger protections than the ASL-2 Standard." |

# Conclusion

Safety frameworks are an important tool and a first line of defence in managing potential risks from frontier AI systems. While they are early-stage and still developing, we hope this paper can help developers who are writing and implementing safety frameworks. We are excited to see this field continue to develop and to see many new safety frameworks presented at the AI Action Summit.

## List of acronyms

AISI    AI Safety Institute (UK)

CCPS  Center for Chemical Process Safety

CISA    Cybersecurity and Infrastructure Agency (US)

DSIT    Department for Science, Innovation, and Technology (UK)

FAISC  Frontier AI Safety Commitments

IEC     International Electrotechnical Commission

IIA     The Institute of Internal Auditors

ISO     International Organization for Standardization

METR  Model Evaluation & Threat Research

NSCS  National Cyber Security Centre (UK)

NIST   National Institute for Standards and Technology (US)

WHO  World Health Organization